\documentclass[12pt,a4paper]{article}

\def\ben{\begin{enumerate}} \def\een{\end{enumerate}}
\def\beq{\begin{equation}} \def\eeq{\end{equation}}
\def\beqn{\begin{equation*}} \def\eeqn{\end{equation*}}
\def\bea{\begin{eqnarray}} \def\eea{\end{eqnarray}}
\def\ba{\begin{array}} \def\ea{\end{array}}
\def\beann{\begin{eqnarray*}} \def\eeann{\end{eqnarray*}}
\def\beasn{\begin{sneqnarray}} \def\eeasn{\end{sneqnarray}}
\def\bi{\begin{itemize}} \def\ei{\end{itemize}}
\def\be{\begin{enumerate}} \def\ee{\end{enumerate}}

\def\ea{\'e}

\def\mf{\mathfrak}

\usepackage{graphicx}
\usepackage{xcolor}
\usepackage[paper=a4paper,left=30mm,right=30mm,top=30mm,bottom=30mm]{geometry}
\usepackage{amsmath}
\usepackage{amsthm}
\usepackage{amssymb}
\usepackage{eufrak}
\numberwithin{equation}{section}

\title{A new approach to classical Einstein-Yang-Mills Theory}

\author
{Donald Salisbury\\
\\
\normalsize{Austin College, 900 North Grand Ave, Sherman, Texas 75090, USA}\\
\normalsize{dsalisbury@austincollege.edu}}

\date{ }

\begin{document}

\maketitle

\begin{abstract}
The conventional Rosenfeld-Bergmann-Dirac constrained Hamiltonian algorithm applied to Einstein-Yang-Mills theory is shown to be equivalent to a local gauge theoretic extension of Cartan's invariant integral approach to classical mechanics. In addition, the Hamiltonian generators of Legendre-projectable spacetime diffeomorphism and gauge symmetries are derived directly as vanishing Noether charges. This leads directly to their interpretation as delivering the correct symmetry variations of both configuration and momentum field variables.
\end{abstract}

\section{Introduction}

I still vividly recall learning about Yang-Mills gauge theory in numerous visits with Bal in the 1970s on Clarendon Avenue in Syracuse. This coupled with his introductory class in group theory was instrumental in my professional preparation. But above all I want to express my gratitude to him in helping me understand the origins and vicissitudes of particle theoretic dual resonance models. I have long identified him in this respect as an informal secondary thesis advisor. I like to characterize my resulting thesis publication in which I constructed a fully relativistic free quantized string in four spacetime dimensions as one of the most widely uncited publications in the physics literature. And I've just confirmed - much to my satisfaction -  that Bal has distinguished himself with a unique reference to this paper! This work was my initial step in a career focus on the Hamiltonian analysis of general covariance, and in this contribution I will present two new approaches to this topic applied to fully covariant general relativity with a Yang-Mills field source.

In part one I will illustrate an alternative derivation of Hamiltonian equations based on an extension of \'Elie Cartan's invariant integral principle to gauge field theory. The method actually constitutes an alternative to the Rosenfeld-Bergmann-Dirac constrained Hamiltonian algorithm. In part two I will present a new derivation of the gauge generators of this model, based on a refinement of the vanishing Noether charge that follows from the underlying general coordinate covariance. I show how the spacetime transformations must be altered and expanded to include not only metric dependence spacetime diffeomorphisms but also related Yang-Mills gauge variations. These results follow from the demand that the generators be projectable under the Legendre transformation from the tangent to the contangent bundle, i.e., from configuration-velocity space to phase space. This work actually constitutes a generalization of the vacuum general relativistic model that is presented in \cite{Salisbury:2022aa}. It turns out that metric dependent spacetime diffeomorphism symmetries cannot by themselves be implemented as phase space transformations. Sundermeyer and I actually showed this long ago \cite{Salisbury:1983ab}. More recently I and my collaborators extended these results to a larger phase space that incorporated the arbitrary gauge variables \cite{Pons:2000aa}, but we invoked a different procedure than that presented in this paper.

\section{An extension Cartan's invariant integral principle}

\'Elie Cartan showed in 1922 that mechanical dynamical equations could be derive from a new invariance principle \cite{Cartan:1922aa}. He supposed that positions $q^i(t)$ and velocities $v^i(t)$ could be introduced as independent variables, and they in addition to the time itself could be taken to be functions of an independent parameter $s$ such that as the parameter value extended say from zero to one, the variables returned to their original value. So, for example, $q^i(t(s);s)\left.\right|_{s=0} = q^i(t(s);s)\left.\right|_{s=1}$. He then formed a closed integral over $s$, and demanded that the integral be independent of $t$. Several examples of an extension of this principle to gauge field theories were given in \cite{Salisbury:2022aa}. Here I extend the idea to the Einstein-Yang-Mills model.

I begin with the Lagrangian 
\beq
\mf{L}_{GRYM} = N \sqrt{g}\left({}^3\!R + K_{ab}K^{ab} -\left(K^a{}_a\right)^2 \right)-{1\over4}N \sqrt{g}
        F^{i}_{\mu\nu}F^{j}_{\alpha\beta}
            g^{\mu\alpha}g^{\nu\beta}C_{ij}. \label{GRYM}
\eeq
The first term is the usual ADM Lagrangian and I employ a version of the Yang-Mills tensor in which the time derivative of the field $A^i_a$ is replaced by an independent function ${\cal V}^{i}$ so that
\beq
F^{i}_{0 a} = {\cal V}^{i}_a - A^i_{0,a} 
        -C^{i}_{jk}A^{j}_{0}A^{k}_{a},
\eeq
and
\beq
F^{i}_{a b} = A^{i}_{b, a}-A^{i}_{a, b}
        -C^{i}_{jk}A^{j}_{a}A^{k}_{b} 
\eeq
The $C^{i}_{jk}$
are the structure constants of the Yang-Mills gauge group and $C_{ij}$ is a nonsingular, symmetric group metric.  (In a semi-simple
group, $C_{ij}$ is usually taken to be $C^{s}_{it}C^{t}_{js}$; in an
Abelian group, one usually takes $C_{ij}=\delta_{ij}$.)

The three-metric time derivatives are replaced by independent fields $v_{ab}$, and inserting these expressions into the canonical momenta $p^{cd} =  \sqrt{g} \left( K^{ab} - K^c{}_c e^{ab}\right)$, with indices raised by the inverse spatial metric $e^{ab}$,  and the $p^{cd}$ now taken to be functions of the $v_{ab}$. I also replace what will become the time derivatives of the lapse and shift $N^\mu$ by the fields $V^\mu$. 

I first carry out the equivalent of a Legendre transformation yielding
\beq
p^{ab} = \frac{\partial \mf{L}_{GRYM}}{\partial v_{ab}} = \sqrt{g} \left(K^{ab} - K^c{}_c e^{ab}\right),
\eeq
where
\beq
K_{ab} = \frac{1}{2N}\left( v_{ab} - N_{a|b} - N_{b|a}\right), \label{K}
\eeq
where indices are raised by $e^{ab}$, the inverse of $g_{ab}$. In addition we have
\beq
{\cal P}^a_i = \frac{\partial \mf{L}_{GRYM}}{\partial {\cal V}^i_a } = \sqrt{g}\left({\cal V}^i_b - A^i_{0,b} -C^i_{jk} A^j_0 A^k_b\right) N^{-1}e^{ab}.
\eeq
I also assume I have the primary constraints
\beq
P_\mu = 0,
\eeq
and
\beq
{\cal P}^0_i = 0,
\eeq
which are respectively the momenta conjugate to $N^\mu$ and $A^i_0$.

The canonical Hamiltonian, expressed in terms of the independent functions $v_{ab}$ and ${\cal V}^i_a $ takes the form
\bea
&&{\cal H}_{GRYM} = p^{ab} v_{ab} + {\cal P}^a_i  {\cal V}^i_a  - \mf{L}_{GRYM} \nonumber \\
&=& \frac{N}{\sqrt{g}} \left( p_{ab} p^{ab} - \left(p^a{}_a\right) \right) - N \sqrt{g} {}^3\!R + 2 N^a_{|b} p^b{}_{a} \nonumber \\
&+& {N\over2\sqrt{g}}C^{ij}g_{ab}{\cal P}^a_i {\cal P}^b_j 
        + N^{a}{\cal P}^b_i F^{i}_{ab}
+{N\sqrt{g}\over4}C_{ij}e^{ac}e^{bd}F^{i}_{ab}F^{j}_{cd}
        + {\cal D}_{a}A^{i}_{0}{\cal P}^a_i  
\eea

Now finally I impose the generalization of Cartan's invariant integral. I require that the closed integral over $s$ as it ranges from zero to one be independent of $t$, i.e.,
\beq
I_{GRYM} = \oint d^3\!x \left( p^{ab} dg_{ab} + P_\mu dN^\mu +{\cal P}^a_i dA^i_a - {\cal H}_{GRYM} dt - P_\mu V^\mu dt- {\cal P}^0_i {\cal V}^i_0 dt\right),  \label{IGRYM}
\eeq
where $dg_{ab} = \frac{dg_{ab}}{ds}ds$, $dN = \frac{dN}{ds} ds$, $dA^i_i =\frac{dA^i_i}{ds} ds$, and $dt = \frac{dt}{ds}ds$ is required to be invariant under independent $\delta$ variations of the field variables. 
I have
\bea
&&0 = \delta I_{GRYM} = \oint d^3\!x \left[\delta p^{ab} dg_{ab} - \delta g_{ab} dp^{ab} + \delta P_\mu dn^\mu - \delta N^\mu dP_\mu +\delta {\cal P}^a_i dA^i_a- \delta A^i_ad{\cal P}^a_i \right. \nonumber \\
&-& \left(\frac{\delta {\cal H}_{GRYM}}{\delta g_{ab}} \delta g_{ab}+ \frac{\delta {\cal H}_{GRYM}}{\delta N^\mu} \delta N^\mu + \frac{\delta {\cal H}_{GRYM}}{\delta p^{ab}} \delta p^{ab}+ \frac{\delta {\cal H}_{GRYM}}{\delta A^i_\mu} \delta A^i_\mu + \frac{\delta {\cal H}_{GRYM}}{\delta {\cal P}^a_i } \delta {\cal P}^a_i \right)dt \nonumber \\
&+& \left(\frac{\delta {\cal H}_{ADM}}{d g_{ab}} d g_{ab}+ \frac{\delta {\cal H}_{ADM}}{\delta N^\mu} d N^\mu + \frac{\delta {\cal H}_{ADM}}{\delta p^{ab}} d p^{ab}+ \frac{\delta {\cal H}_{GRYM}}{\delta A^i_\mu} d A^i_\mu + \frac{\delta {\cal H}_{GRYM}}{\delta {\cal P}^a_i } d {\cal P}^a_i\right)\delta t \nonumber \\
&-&\left. \delta P_\mu V^\mu dt - \delta V^\mu P_\mu dt + dP_\mu V^\mu \delta t + P_\mu dV^\mu \delta t  \right. \nonumber \\
&-& \left. \delta {\cal P}^0_i {\cal V}^i_0 dt-\delta {\cal V}^i_0 {\cal P}^0_i  dt+d{\cal P}^0_i {\cal V}^i_0 \delta t + {\cal P}^0_i d{\cal V}^i_0 \delta t\right].
\eea
From the required vanishing of the coefficient of $\delta p^{ab}$ we conclude that $\frac{\partial g_{ab}}{\partial t} = \frac{\delta {\cal H}_{GRYM}}{\delta p^{ab}} $, from $\delta g_{ab}$ that $\frac{\partial p^{ab}}{\partial t}= -\frac{\delta {\cal H}_{GRYM}}{\delta g_{ab}} $, from $\delta P_\mu$ that $\frac{\partial N^\mu}{ \partial t} = V^\mu$,  , from $\delta {\cal P}^a_i$ that $\frac{\partial A^i_a}{\partial t} = \frac{\delta {\cal H}_{GRYM}}{\delta {\cal P}^a_i} $, from $\delta A^i_a$ that $\frac{\partial {\cal P}^a_i}{\partial t} = - \frac{\delta {\cal H}_{GRYM}}{\delta A^i_a}$, from $\delta {\cal P}^0_i$ that $\frac{\partial A^i_0}{\partial t} = {\cal V}^i_0$, from $\delta N^\mu$ that $\frac{\partial P_\mu}{ \partial t} = - \frac{\delta {\cal H}_{ADM}}{\delta N^\mu}$,  and from $\delta A^i_0$ that $\frac{\partial {\cal P}^0_i}{\partial t} = - \frac{\delta {\cal H}_{GRYM}}{\delta A^i_0} $. Since the primary constraint must be conserved these latter two relations gives us the secondary constraints
\beq
{\cal H}_0 :=  \frac{1}{\sqrt{g}} \left(p_{ab}p^{ab} - (p^c{}_c)^2) \sqrt{g} {}^3\!R \right)+{1\over2\sqrt{g}}C^{ij}g_{ab}{\cal P}^a_i {\cal P}^b_j 
+{\sqrt{g}\over4}C_{ij}e^{ac}e^{bd}F^{i}_{ab}F^{j}_{cd} = 0, \label{H}
\eeq
\beq
 {\cal H}_a := - 2 p^b{}_{a|b}+{\cal P}^b_i F^{i}_{ab} = 0,
\eeq
and
\beq
{\cal G}_i :=- {\cal D}_{a}{\cal P}^a_i  = 0.
\eeq
With these results it turns out that the coefficient of $\delta t$ is zero as required. Thus the invariant integral approach has delivered the known Hamiltonian analysis of general relativity with a Yang-Mills field source.

\section{The generator of diffeomorphism-induced plus gauge phase space transformations}

I next present a new derivation of the generators of Legendre-projectable symmetry transformations. The procedure follows closely the second Noether theorem-based approach pioneered by L\'eon Rosenfeld in 1930 \cite{Rosenfeld:1930aa} .\footnote{See \cite{Rosenfeld:2017ab} for a translation into English of this article by myself and Kurt Sundermeyer and \cite{Salisbury:2017aa} for a careful analysis of the article and its relation to later work.}  Rosenfeld however did not address the projectability problem. But it is in recognition of Rosenfeld's generally unappreciated development of constrained Hamiltonian dynamics that I and my co-authors have identified the following procedure as the Rosenfeld-Bergmann-Dirac method.

I consider  transformed solutions of the Einstein Yang Mills equations obtained through an infinitesimal active coordinate transformation $x'^\mu = x^\mu - \epsilon^\mu(x)$.  Rather than simply displace the new solutions to $x^\mu - \epsilon^\mu(x)$, I map the old solutions to this new location via the active manifold map. Note that from this active perspective the coordinates are not altered, rather, the solutions as functions of these coordinates are shifted. The variation under these circumstances is simply the Lie derivative with respect to $\epsilon^\mu$ - and continuing the tradition that began with Noether, I will represent these variations by $\bar \delta$.\footnote{See \cite{Salisbury:2020aa} for Bergmann's use of Noether's notation. Note also that this is a special case of the variations that were represented by $\delta_0$ in \cite{Salisbury:2022ab} .} I obtain
\beq
 \bar \delta g_{ab} = 2 g_{c (b}\epsilon^c_{,a)} + 2 g_{c (a}N^c \epsilon^0_{,b)} +\dot g_{ab}\epsilon^0 + g_{ab,c} \epsilon^c, \label{dgab}
\eeq
\beq
 \bar \delta N  =  N \dot \epsilon^0 - N N^a \epsilon^0_{,a} + \dot N \epsilon^0 + N_{,a} \epsilon^a, \label{dn}
\eeq
\beq
 \bar \delta N^a =  N^{a}\dot\epsilon^{0}
-(N^{2}e^{ab}+N^{a}N^{b})\epsilon^{0}_{,b} + \dot \epsilon^a - N^b \epsilon^a_{,b} +\dot N^{a}\epsilon^{0} + N^{a}_{,b}\epsilon^b. \label{dna}
\eeq
and
\beq
\bar \delta A^i_\mu = A^i_\nu \epsilon^\nu_{,\mu} + A^i_{\mu,\nu}\epsilon^\nu. \label{da}
\eeq

Before inserting these variations in the action I will first derive the corresponding vanishing charge that follows from Noether's second theorem. 
Noether's second theorem is applicable in this case since the action is invariant under the active diffeomorphism-induced field transformations. This is a consequence of  the fact that Lagrangian transforms as a scalar density under these transformations - excepting for variations at spatial infinity which can be taken to vanish. Consequently, we have
\beq
\bar \delta {\cal L}_{GRYM} \equiv \left({\cal L}_{GRYM}  \epsilon^\mu\right)_{,\mu}. \label{ident}
\eeq
Therefore, given that solutions are transformed into solutions, we have according to (\ref{ident})
\bea
0 &=&   \int d^4x \left[\bar \delta  {\cal L}_{GRYM} - \left({\cal L}_{GRYM} \epsilon^\mu\right)_{,\mu}\right] \nonumber \\
&=&  \int d^4x \left(\frac{\partial {\cal L}_{GRYM} }{\partial g_{ab,\mu}} \bar \delta g_{ab} + \frac{\partial {\cal L}_{GRYM} }{\partial  N^\nu_{,\mu}} \bar \delta N^\nu + \frac{\partial {\cal L}_{GRYM} }{\partial  A^i_{\nu,\mu}} \bar \delta A^i_\nu- {\cal L}_{GRYM} \epsilon^\mu\right)_{, \mu} \nonumber \\
&=&\left. \int d^3x\left(\frac{\partial {\cal L}_{ADM} }{\partial g_{ab,0}} \bar \delta g_{ab} + \frac{\partial {\cal L}_{ADM} }{\partial  N^\nu_{,0}} \bar \delta N^\nu +\frac{\partial {\cal L}_{GRYM} }{\partial  A^i_{\nu,0}} \bar \delta A^i_\mu- {\cal L}_{GRYM} \epsilon^0\right)\right|_{x^0_i}^{x^0_f} \nonumber \\
&=& \left. \int d^3x\left(p^{ab} \bar \delta g_{ab} + {\cal P}^a_i \bar \delta A^i_a + P_\mu \bar \delta N^\nu + {\cal P}^0_i \bar \delta A^i_0- {\cal L}_{GRYM} \epsilon^0\right)\right|_{x^0_i}^{x^0_f} \label{Noether}
\eea
It is noteworthy that Rosenfeld actually derived the equivalent conserved quantity for a general relativistic dynamical tetrad field in interaction with electrodynamic and spinorial fields.

There is however a problem. It is not projectable to phase space because of the appearance of the time derivatives $\dot N^\mu$ and $\dot A^i_0$ as we note in substituting the variations (\ref{dgab}) through (\ref{da}). Representing what we have called the Rosenfeld-Noether vanishing charge density by $\mf{C}^0_{GRYM} $, we have
\bea
&&0 =\int d^3x\,\mf{C}^0_{GRYM} \nonumber \\
&& =  \int d^3x \left[ p^{ab} \left(2 g_{c (b}\epsilon^c_{,a)} + 2 g_{c (a}N^c \epsilon^0_{,b)} +\dot g_{ab}\epsilon^0 + g_{ab,c}  \epsilon^c\right) \right.\nonumber \\
&+& {\cal P}^a_i \left( A^i_\nu \epsilon^\nu_{,a} + A^i_{a,b}\epsilon^b+ \dot A^i_a \epsilon^0\right)- {\cal L}_{GRYM} \epsilon^0  \label{1}\\
&+& P_0 \left(N \dot \epsilon^0 - N N^a \epsilon^0_{,a} + \dot N \epsilon^0 + N_{,a} \epsilon^a\right) \label{2} \\
&+& P_a \left(N^{a}\dot\epsilon^{0}
-(N^{2}e^{ab}+N^{a}N^{b})\epsilon^{0}_{,b} + \dot \epsilon^a - N^b \epsilon^a_{,b} +\dot N^{a}\epsilon^{0} + N^{a}_{,b}\epsilon^b\right)\label{3} \\
&&+ \left.{\cal P}^0_i  \left( A^i_\nu \dot \epsilon^\nu + \dot A^i_0\epsilon^0 + A^i_{0,a}\epsilon^a \right)\right] \label{4}
\eea
I concentrate first on (\ref{2}) and (\ref{3}) where the unprojectable time derivatives of the lapse and shift fields appear. As was shown in \cite{Pons:1997aa}, these can and must be eliminated through gravitational field dependent infinitesimal transformations $\epsilon^\mu = n^\mu \xi^0 + \delta^\mu_a \xi^a$ where $n^\mu = \left(N^{-1}, - N^{-1} N^a \right)$ is the orthonormal to the constant time surfaces. It follows that
\beq
\dot \epsilon^0 = - N^{-2} \dot N \xi^0+ N^{-1} \dot \xi^0, \label{dot0}
\eeq
\beq
\epsilon^0_{,a} = - N^{-2} N_{,a} \xi^0 + N^{-1} \xi^0_{,a},
\eeq
\beq
\dot \epsilon^a = N^{-2} \dot N N^a \xi^0 - N^{-1} \dot N^a \xi^0 - N^{-1}  N^a \dot \xi^0 + \dot \xi^a, \label{dota}
\eeq
and
\beq
\epsilon^a_{,b} = N^{-2} N_{,b} N^a \xi^0- N^{-1} N^a_{,b} \xi^0 - N^{-1} N^a \xi^0_{,b} + \xi^a_{,b}.
\eeq
Substitution into (\ref{2}) and (\ref{3}) yields
\beq
P_0 \left(\dot \xi^0 - N^a \xi^0_{,a} + N_{,a} \xi^a\right),
\eeq
and
\beq
P_a \left(N_{,b} e^{ab} \xi^0 - N e^{ab} \xi^0_{,b} + \dot \xi^a - N^b \xi^a_{,b} + N^a_{,b} \xi^b\right).
\eeq

Next, substituting $\epsilon^\mu = n^\mu \xi^0 + \delta^\mu_a \xi^a$ into (\ref{1}) I obtain first 
\beq
\left(p^{ab} \dot g_{ab} + {\cal P}^a_i \dot A^i_a - \mf{L}_{GRYM}\right) \epsilon^0 = {\cal H}_0 \xi^0,
\eeq
where ${\cal H}_0$ is the secondary constraint given by 
 (\ref{H}). 
The remaining terms in (\ref{1}) are
\bea
&&-2 p^{ab} g_{cb}N^c_{,a}  N^{-1} \xi^0 - p^{ab} g_{ab,c} N^c N^{-1} \xi^0 + 2 p^{ab} g_{cb} \xi^c_{,a} + p^{ab} g_{ab,c} \xi^c + 2 p^{ab}N_{a|b} N^{-1} \xi^0 \nonumber \\
&& 
\eea

But we still have in addition to the unprojectable $\dot A^i_0$ term in (\ref{4}) the return of time derivatives of the lapse and shift that result from the time derivative of $\epsilon^\mu$. Fortunately there is a way of eliminating all of these unprojectable time derivatives by making use of the additional Yang-Mills local gauge symmetry of the form
\beq
\delta_\Lambda A^i_\mu = - \Lambda^i_{,\mu} - C^i_{jk} \Lambda^j A^k_\mu.
\eeq

The action is invariant under this transformation, and there is therefore a corresponding vanishing Noether charge density which can be derived in a manner similar to (\ref{Noether}), namely, this equation simplifies in this case to
\beq
0 = \left. \int d^3x\left({\cal P}^a_i \bar \delta_\Lambda A^i_a  + {\cal P}^0_i \bar \delta_\Lambda A^i_0 \right) \right|_{x^0_i}^{x^0_f}.
\eeq
Then because the time dependence of $\Lambda^i$ is arbitrary we deduce the existence of the vanishing Noether charge density
\beq
\mf{C}_{\Lambda} = -{\cal P}^\mu_i  \left(  \Lambda^i_{,\mu} + C^i_{jk} \Lambda^j A^k_\mu\right). 
\eeq
Fortunately, the form of the diffeomorphism variation given in (\ref{da}) immediately suggests an appropriate additional gauge transformation that will eliminate the time derivatives of $\epsilon^\nu$. We add a gauge variation with $\Lambda = A^i_\nu \epsilon^\nu =A^i_\nu n^\nu \xi^0 + A^i_a \xi^a = : {}^\xi\!\Lambda$ which delivers a net variation
\beq
\delta A^i_\mu := \bar \delta A^i_\mu+ \delta_{{}^\xi\!\Lambda} A^i_\mu =  F^i_{\nu \mu} \epsilon^\nu,
\eeq
i.e.,
\bea
\delta A^i_0 &=& -F_{0a}\left(-N^{-1} N^a \xi^0 + \xi^a\right) \nonumber \\
&=& C^{ij} g_{ab} {\cal P}^b_j N^a \xi^0 + \left(N C^{ij} g_{ab} {\cal P}^b_j + N^b F^i_{ba}\right) \xi^a \label{a0}
\eea 
and
\bea
\delta A^i_a &=& F^i_{0a} N^{-1} \xi^0 - F_{ab} \left(- N^{-1} N^b \xi^0 + \xi^b\right) \nonumber \\
&=& \left(\dot A^i_a -{\cal D}_a A^i_0 \right) \xi^0 - F_{ab}^i \xi^b, \label{aa}
\eea
where in (\ref{a0})  I used the result from the field equations that $F^i_{0a} = N C^{ij} g_{ab} {\cal P}^b_j + N^b F^i_{ba}$, and in (\ref{aa}) the fact that it is also true that $F^i_{0a} = \dot A^i_a - {\cal D}_a A^i_0$.

Returning to the vanishing integrand of (\ref{Noether}) I replace the $\bar \delta$ variations by the projectable $\delta$ variations, thereby delivering the vanishing Legendre projectable Rosenfeld Noether charge density. I omit some calculation details here dealing with the metric field variations. These computations can be found in  \cite{Salisbury:2022aa}. I obtain the vanishing generator
\bea
&&{\cal C}(\xi) = p^{ab} \delta g_{ab} + {\cal P}^a_i \delta A^i_a + P_\mu \delta  N^\nu + {\cal P}^0_i  \delta A^i_0- {\cal L}_{GRYM} N^{-1} \xi^0 \nonumber \\
&&= {\cal H}_0 \xi^0 -2 p^{ab} g_{cb}N^c_{,a}  N^{-1} \xi^0 - p^{ab} g_{ab,c} N^c N^{-1} \xi^0 + 2 p^{ab} g_{cb} \xi^c_{,a} + p^{ab} g_{ab,c} \xi^c + 2 p^{ab}N_{a|b} N^{-1} \xi^0 \nonumber \\
&& \nonumber \\
&&+P_0 \left(\dot \xi^0 - N^a \xi^0_{,a} + N_{,a} \xi^a\right)+
P_a \left(N_{,b} e^{ab} \xi^0 - N e^{ab} \xi^0_{,b} + \dot \xi^a - N^b \xi^a_{,b} + N^a_{,b} \xi^b\right) \nonumber \\ \nonumber \\
&&- {\cal P}^a_i \left( {\cal D}_a A^i_0  \xi^0 + F_{ab}^i \xi^b\right) \nonumber \\
&&+ {\cal P}^0_i\left( C^{ij} g_{ab} {\cal P}^b_j N^a \xi^0 + \left(N C^{ij} g_{ab} {\cal P}^b_j + N^b F^i_{ba}\right) \xi^a \right)
\eea

This is not entiely equivalent to the generator that was obtained in \cite{Pons:2000aa} through a different route, and that group-theoretic calculation needs to be revisited. There is here, however, no doubt that this is the correct generator. It is based directly on the known transformation rules for the configuration field variables. And indeed, given that it relies directly on the phase space one-form $p_a dq^a$ it is clear that it also generates the correct variations of the momentum variables. This was actually proven already in 1930 by Rosenfeld - but only for the Legendre-projectable case.

\bibliographystyle{apalike}
\bibliography{qgrav-V19}

\end{document}